\documentclass{ws-procs961x669}
\usepackage[mathscr]{euscript}
\usepackage[figuresright]{rotating}
\usepackage{array,multirow,makecell} 
\usepackage[breaklinks, colorlinks, citecolor=blue]{hyperref}
\usepackage{url,doi}
\usepackage{float}
\usepackage{graphicx}
\usepackage{epstopdf}
\usepackage{bm}
\usepackage{booktabs}

\begin{document}

\def\dzero{{(0)}}
\def\dtwo{{\prime\prime}}
\def\dthree{{\prime\prime\prime}}
\def\dfour{{(4)}}
\def\dfive{{(5)}}
\def\dsix{{(6)}}
\def\dn{{(n)}}

\def\lsim{\,\lower2truept\hbox{${< \atop\hbox{\raise4truept\hbox{$\sim$}}}$}\,}
\def\gsim{\,\lower2truept\hbox{${> \atop\hbox{\raise4truept\hbox{$\sim$}}}$}\,}

%\wstoc{For proceedings contributors: Using World Scientific's\\ WS-procs961x669 document class in \LaTeX2e}
%{A. B. Author and C. D. Author}

\title{ Observer motion and boosting effects on the cosmic background monopole spectrum, solutions and perspectives}

\author{T. Trombetti$^*$}
%\index{author}{Author, A. B.} % or 
%\aindx{Author, A. B.}

\address{INAF, Istituto di Radioastronomia\\
Via Piero Gobetti, 101, 40129 Bologna, Italy\\
$^*$E-mail: trombetti@ira.inaf.it}

\begin{abstract}
The peculiar motion of an observer relative to an ideal reference frame at rest with respect to the cosmic background produces boosting effects which modify and transfer at higher multipoles the frequency spectrum of the isotropic background. To mitigate the computational effort needed for accurate theoretical predictions, analytical solutions of a linear system able to evaluate the spherical harmonic expansion coefficients for (analytical or semi-analytical) background representations have been presented, and extended to generic tabulated functions potentially affected by numerical uncertainties. Owing to the dipole spectrum frequency dependence and to precise inter-frequency calibrations, it will be possible to constrain (or even detect) the tiny imprints in the background spectrum from a variety of cosmological and astrophysical processes.
\end{abstract}

% Keywords with PACS numbers  https://ufn.ru/en/pacs/

\bodymatter

\section{Introduction}

In presence of a peculiar motion of the observer, the frequency spectrum of the background isotropic monopole emission is modified and transferred to higher multipoles by boosting effects. The modified spectrum can be investigated through the frequency dependence of the signal variation in the sky using the complete description of the Compton-Getting effect. The largest effect is on the dipole at $\ell = 1$, mainly attributed to the motion of the solar system barycentre (Sect.\,\ref{sec:theoframe}).

Expanding the monopole frequency spectrum in Taylor's series and starting from the analytical solutions for the spherical harmonic coefficients, it is possible to derive explicit expressions for the harmonic coefficients involving monopole frequency spectrum derivatives and their scaling relations. For generic background functions, a method for fast computing the coefficients of the spherical harmonic expansion up to the multipole $\ell_{max} = 6$ is presented (Sect.\,\ref{sec:FES}).

Usually, whether the monopole spectrum can be well represented by analytical or semi-analytical functions and the computation is supplied by the required level of numerical accuracy, numerical instabilities do not affect the frequency spectra at high multipoles. When dealing with tabulated functions typically characterized by a poor frequency resolution compared to the frequency Doppler shift ($\delta \nu / \nu \sim \beta = {\rm v} / c$, ${\rm v}$ being the observer speed), an interpolation algorithm is required. Moreover, tabulated functions could be affected by numerical uncertainties connected to intrinsic inaccuracies in astrophysical modelling or in related ingested observations or to limited accuracy in their numerical computation. The propagation of these uncertainties can preclude, in principle, an accurate computation of the monopole transfer at any multipole.
For this reason, we implemented different low-pass filters: pre-filtering in Fourier space of the tabular representation; filtering in real and Fourier space of the numerical derivatives; interpolation approaches; a dedicated method based on the boosted signal amplification and deamplification (Sect.\,\ref{sec:filt}). 
The quality of these methods for suitable analytical approximations of the tabulated functions, possibly polluted with simulated noise, is evaluated. I discuss the improvement achievable in the results for some specific background spectra (Sect.\,\ref{sec:res}). Finally, I describe the potential advance with respect to the Far Infrared Absolute Spectrophotometer (FIRAS) in the recovery of some characteristic distortion parameters from dipole analyses applied to nearly all-sky surveys, as well as the perspectives in background spectrum reconstruction with high resolution and sensitivity observations of sky areas of a few square degrees (Sect.\,\ref{sec:obsperspe}).

\section{Theoretical framework}
\label{sec:theoframe}

The Compton-Getting effect \citep{1970PSS1825F} is based on the Lorentz invariance of the photon distribution function, $\eta(\nu)$; it allows to describe how the peculiar velocity of an observer impacts the background frequency spectrum. At the frequency $\nu$, the observed signal in equivalent thermodynamic temperature, $T_{\rm th} (\nu) ={(h\nu/k)} / {\ln(1+1/\eta(\nu))}$, is

\begin{align}\label{eq:eta_boost}
T_{\rm th}^{\rm BB/dist} (\nu, {\hat{n}}, \vec{\beta}) =
\frac{xT_{0}} {{\rm{ln}}(1 + 1 / (\eta(\nu, {\hat{n}}, \vec{\beta}))^{\rm BB/dist}) } = \frac{xT_{0}} {{\rm{ln}}(1 + 1 / \eta(\nu')) }\, ,
\end{align}

\noindent where $k$ and $h$ are the Boltzmann and Planck constants and $\eta(\nu, {\hat{n}}, \vec{\beta}) = \eta(\nu')$, with

\begin{equation}\label{eq:nuboost}
\nu' = \nu (1 - {\hat{n}} \cdot \vec{\beta})/(1 - \beta^2)^{1/2} \, .
\end{equation}
\noindent
In the above formulas, ${\hat{n}}$ is the sky direction unit vector, $\vec{\beta} = \vec{{\rm v}} / c$ is the dimensionless observer velocity, $x=h\nu/(kT_r)$ is the cosmic microwave background (CMB) redshift invariant dimensionless frequency, $T_r=T_0(1+z)$ is the CMB redshift dependent effective temperature, and $T_0$ is the current CMB effective temperature in the blackbody spectrum approximation such that $aT_0^4$ gives the current CMB energy density with $a = 8\pi I_3 k^4 / (hc)^3$, $I_3 = \pi^4/15$. The joint analysis of the data from FIRAS and from the Wilkinson Microwave Anisotropy Probe (WMAP), gives $T_0 = (2.72548 \pm 0.00057)$\,K.\cite{2009ApJ...707..916F} The notation `BB/dist' indicates a blackbody spectrum or any type of spectral distortion.\citep{2018JCAP...04..021B}

\section{Fully explicit solutions}
\label{sec:FES}

Typically, the solutions derived through Eq.\,\eqref{eq:eta_boost} are formally exact but, for a given observer velocity, the effect need to be estimated for all the relevant directions, making this approach extremely time consuming, particularly at high resolution. A different approach consists in expanding Eq.\,\eqref{eq:eta_boost} in spherical harmonics

\begin{equation}\label{eq:harm}
T_{\rm th}^{\rm BB/dist} (\nu, \theta, \phi, \beta) =
\sum_{\ell=0}^{\ell_{\rm max}} \sum_{m=-\ell}^{\ell} a_{\ell,m} (\nu, \beta) Y_{\ell,m}(\theta, \phi) \, 
\end{equation}

\noindent to be inverted to derive the $a_{\ell,m} (\nu, \beta)$ coefficients

\begin{align}\label{eq:harminv}
& a_{\ell,m} (\nu, \beta) = \int_{\theta=0}^\pi \int_{\phi=0}^{2\pi} T_{\rm th}^{\rm BB/dist} (\nu, \theta, \phi, \beta) e^{-im\phi} {\tilde P}_\ell^m ({\rm cos} \, \theta) \, {\rm sin} \, \theta \, d\theta \, d\phi \, \nonumber
\\ & = \int_{\theta=0}^\pi \int_{\phi=0}^{2\pi} \left[{ T_{\rm th}^{\rm BB/dist} (\nu, \theta, \phi, \beta) - T_{\rm th}^{\rm BB/dist,rest} (\nu) }\right] 
\\ & \cdot e^{-im\phi} {\tilde P}_\ell^m ({\rm cos} \, \theta) \, {\rm sin} \, \theta \, d\theta \, d\phi 
+ a_{\ell,m}^{\rm rest} (\nu) \, , \nonumber
\end{align}

\noindent where $Y_{\ell,m}$ reduce to the renormalised associated Legendre's polynomials ${\tilde P}_\ell^m$.
Remarkably, choosing a reference system with the $z$-axis parallel to the observer velocity allows to simplify the problem, since it keeps the dependence on the colatitude, $\theta$, while the one on the longitude, $\phi$, vanishes. In this way, only the spherical coefficients $a_{\ell,m} (\nu, \beta)$ with $m=0$ do not vanish.

The last equality in Eq.\,\eqref{eq:harminv} is useful in the numerical computation because the integrand function becomes the difference between the equivalent thermodynamic temperatures in the reference frames in motion and at rest with respect to the background. In general, this approach requires a delicate and computationally demanding integration over $\theta$. Since $\beta$ is small, it could be difficult to achieve the extreme precision needed to characterize the very fine details of spectral features.

\subsection{Solutions in terms of linear combinations}

Alternatively, we can construct a linear system of equations in the $N$ unknowns $a_{\ell,0} (\nu,\beta)$, with $\ell=0,N-1$, corresponding to Eq.\,\eqref{eq:harm} with $m=0$ and different colatitudes $\theta_i$ with, again, $i=0,N-1$. Since the determinant of the coefficient matrix does not vanish, the system can be solved as linear combinations of these  $N$ signals $T_{\rm th}^{\rm BB/dist} (\nu, \theta_i, \phi, \beta)$. 

Being $\beta \simeq 1.2336 \times10^{-3}$, \citep{2020A&A...641A...1P} adopting $\ell_{\rm max}=6$ in Eq.\,\eqref{eq:harm}, or, equivalently,
computing the signal through Eq.\,\eqref{eq:eta_boost} in only $N = \ell_{\rm max}+1$ sky directions will allow us to achieve a high numerical accuracy, suitable for any application even in the very far future.

The amplitude of $a_{\ell,m} (\nu, \beta)$ decreases at increasing multipole, $\ell$, as $\beta^{\ell \cdot p}$, with $p \approx 1$ (for a blackbody $p=1$ and $a_{\ell,m} (\nu, \beta)$ does not depend on $\nu$).

For appropriate choices of the $N$ colatitudes $\theta_i$, symmetry properties of the renormalised associated Legendre polynomials with respect to $\pi/2$, allow to divide the system in two subsystems, one for $\ell=0$ and even multipoles and the other for odd multipoles. This separation improves the solution accuracy since, neglecting higher $\ell$'s produces an error dominated by the $\ell_{\rm max}+2$ term for $\ell = 0$ and even $\ell$ (or by $\ell_{\rm max}+1$ for odd $\ell$).\citep{2021A&A...646A..75T}
Similar considerations hold for any $\ell_{\rm max}$. For $\ell_{\rm max}=6$, we select $\theta_i = 0, \pi/4, \pi/3, \pi/2, (2/3)\pi, (3/4)\pi$ and $\pi$ to simplify the algebra. 

Defining $w_i = {\rm cos} \, \theta_i = 1, \sqrt{2}/2, 1/2, 0, -1/2, -\sqrt{2}/2, -1$, the solutions are\citep{2021A&A...646A..75T}

\begin{align}
\label{eq:struct_even}
a_{\ell,0} & = A_\ell \sqrt{\frac{4\pi}{2\ell+1}} \Bigg[ d_{\ell,1} \left({T_{\rm th}^{\rm BB/dist} (w=1) + T_{\rm th}^{\rm BB/dist} (w=-1) }\right) \nonumber
\\ & + d_{\ell,2} \left({T_{\rm th}^{\rm BB/dist} (w=\sqrt{2}/2) + T_{\rm th}^{\rm BB/dist} (w=-\sqrt{2}/2) }\right) \nonumber
\\ & + d_{\ell,3} \left({T_{\rm th}^{\rm BB/dist} (w=1/2) + T_{\rm th}^{\rm BB/dist} (w=-1/2) }\right) \nonumber
\\ & + d_{\ell,4} T_{\rm th}^{\rm BB/dist} (w=0) \Bigg] \,
\end{align}
\noindent
for $\ell=0$ and even multipoles, and 
\begin{align}
\label{eq:struct_odd}
a_{\ell,0} & = A_\ell \sqrt{\frac{4\pi}{2\ell+1}} \Bigg[ d_{\ell,1} \left({T_{\rm th}^{\rm BB/dist} (w=1) - T_{\rm th}^{\rm BB/dist} (w=-1) }\right) \nonumber
\\ & + d_{\ell,2} \left({T_{\rm th}^{\rm BB/dist} (w=\sqrt{2}/2) - T_{\rm th}^{\rm BB/dist} (w=-\sqrt{2}/2) }\right) \nonumber
\\ & + d_{\ell,3} \left({T_{\rm th}^{\rm BB/dist} (w=1/2) - T_{\rm th}^{\rm BB/dist} (w=-1/2) }\right) \Bigg]\,
\end{align}
\noindent
for odd multipoles. For the assumed $\ell_{max}$ and $\theta_i$, Table \ref{tab:coeffs} gives the coefficients $A_\ell$ and $d_{\ell,i}$.\cite{2024AA...684A..82T}

\begin{table}
\tbl{ \small $A_\ell$ and $d_{\ell,i}$ coefficients.}
{\begin{tabular}{@{}c|c|c|c|c|c@{}}
\toprule
      $\ell$ & $A_\ell$ & $d_{\ell,1}$ & $d_{\ell,2}$ & $d_{\ell,3}$ & $d_{\ell,4}$ \\ [0.2ex]
        \hline
0 & 1/630 & 29 & 120 & 64 & 204 \\[0.2ex]
1 & 1/210 & 29 & 60$\sqrt{2}$ & 32 & -- \\[0.2ex]
2 & 1/693 & 121 & 396 & $-$352 & $-$330 \\[0.2ex]
3 & 2/135 & 13 & 15$\sqrt{2}$ & $-$56 & -- \\[0.2ex]
4 & 8/385 & 9 & $-$10 & $-$16& 34\\[0.2ex]
5 & 32/189 & 1 & $-$3$\sqrt{2}$ & 4 & -- \\[0.2ex]
6 & 64/693 & 1 & $-$6 & 8 & $-$6 \\ [0.2ex]
    \bottomrule
    \end{tabular}
}
\label{tab:coeffs}
\end{table}

\subsection{Solutions in terms of derivatives}

The structure of the solutions described by Eqs.\,\eqref{eq:struct_even} and \eqref{eq:struct_odd} presents some similarities with the weights for the centred approximation numerical derivative scheme, \citep{1988MC} as already discussed in Ref.\citenum{2021A&A...646A..75T}. This evidence indicates that there is a close relationship with the derivatives of the considered signal, as underlined for the first time in Ref.\citenum{1981A&A....94L..33D} for the dipole.

The values of $T_{\rm th}^{\rm BB/dist} (w)$ should be always very close to the one computed at $w=0$, i.e. in the direction perpendicular to the observer motion.
Then, if $T_{\rm th}^{\rm BB/dist} (w)$ can be expanded in Taylor's series around $w=0$, by adopting the Lagrange notation and denoting the derivatives of
$T_{\rm th}^{\rm BB/dist} (w)$ performed with respect to $w$, evaluated at $w=0$, with $T_{\rm th}^\dzero$, $T_{\rm th}'$, ..., $T_{\rm th}^\dsix$, from order zero to six,
for the adopted set of $w_i$ and after some calculations, Eqs. \eqref{eq:struct_even} and \eqref{eq:struct_odd} can be rewritten as

\begin{align}
\label{eq:struct_der_even}
a_{\ell,0} & = 2\, A_\ell \sqrt{\frac{4\pi}{2\ell+1}} \Bigg[ \Bigg(d_{\ell,1} + d_{\ell,2} + d_{\ell,3}\Bigg)\, T_{\rm th}^\dzero \nonumber
\\ & + \frac{1}{2!} \,\Bigg(d_{\ell,1} + \frac{1}{2} d_{\ell,2} + \frac{1}{4} d_{\ell,3}\Bigg)\, T_{\rm th}^\dtwo \nonumber
\\ & + \frac{1}{4!} \,\Bigg(d_{\ell,1} + \frac{1}{4} d_{\ell,2} + \frac{1}{16} d_{\ell,3}\Bigg)\, T_{\rm th}^\dfour \nonumber
\\ & + \frac{1}{6!} \,\Bigg(d_{\ell,1} + \frac{1}{8} d_{\ell,2} + \frac{1}{64} d_{\ell,3}\Bigg)\, T_{\rm th}^\dsix + \frac{1}{2}\,d_{\ell,4}\, T_{\rm th}^\dzero \Bigg] \,
\end{align}

\noindent for $\ell=0$ and even multipoles, and

\begin{align}
\label{eq:struct_der_odd}
a_{\ell,0} & = 2\, A_\ell \sqrt{\frac{4\pi}{2\ell+1}} \Bigg[ \Bigg(d_{\ell,1} + \frac{\sqrt{2}}{2}d_{\ell,2} + \frac{1}{2}d_{\ell,3}\Bigg)\, T_{\rm th}' \nonumber
\\ & + \frac{1}{3!} \,\Bigg(d_{\ell,1} + \frac{\sqrt{2}}{4}d_{\ell,2} + \frac{1}{8}d_{\ell,3}\Bigg)\, T_{\rm th}^\dthree \nonumber
\\ & + \frac{1}{5!} \,\Bigg(d_{\ell,1} + \frac{\sqrt{2}}{8}d_{\ell,2} + \frac{1}{32} d_{\ell,3}\Bigg)\, T_{\rm th}^\dfive \Bigg] \,
\end{align}

\noindent for odd multipoles. As a consequence of the separation of the system into two subsystems, from Eqs. \eqref{eq:struct_der_even} and \eqref{eq:struct_der_odd} emerge that only the derivatives of even (odd) order contribute to $a_{\ell,0}$ for even (odd) $\ell$. In conclusion, inserting the values of the $A_\ell$ and $d_{\ell,i}$ coefficients given in Table\,\ref{tab:coeffs}, we have\cite{2024AA...684A..82T}

\begin{equation}
\label{eq:struct_der_0}
a_{0,0} = \sqrt{4\pi} \; \left[T_{\rm th}^\dzero  + \frac{1}{6} T_{\rm th}^\dtwo + \frac{1}{120} T_{\rm th}^\dfour + \frac{1}{5040} T_{\rm th}^\dsix \right] \, ,
\end{equation}
\begin{equation}
\label{eq:struct_der_1}
a_{1,0} = \sqrt{\frac{4\pi}{3}} \; \left[T_{\rm th}'  + \frac{1}{10} T_{\rm th}^\dthree + \frac{1}{280} T_{\rm th}^\dfive \right] \, ,
\end{equation}
\begin{equation}
\label{eq:struct_der_2}
a_{2,0} = \frac{1}{3}\sqrt{\frac{4\pi}{5}} \; \left[T_{\rm th}^\dtwo + \frac{1}{14} T_{\rm th}^\dfour + \frac{1}{504} T_{\rm th}^\dsix \right] \, ,
\end{equation}
\begin{equation}
\label{eq:struct_der_3}
a_{3,0} = \frac{1}{15}\sqrt{\frac{4\pi}{7}} \; \left[T_{\rm th}^\dthree + \frac{1}{18} T_{\rm th}^\dfive \right] \, ,
\end{equation}
\begin{equation}
\label{eq:struct_der_4}
a_{4,0} = \frac{1}{105}\sqrt{\frac{4\pi}{9}} \; \left[T_{\rm th}^\dfour + \frac{1}{22} T_{\rm th}^\dsix \right] \, ,
\end{equation}
\begin{equation}
\label{eq:struct_der_5}
a_{5,0} = \frac{1}{945}\sqrt{\frac{4\pi}{11}} \; T_{\rm th}^\dfive \, ,
\end{equation}
\begin{equation}
\label{eq:struct_der_6}
a_{6,0} = \frac{1}{10395}\sqrt{\frac{4\pi}{13}} \; T_{\rm th}^\dsix\, ,
\end{equation}

\noindent where each denominator, $D_\ell$, in front of the square root can be rewritten as $D_\ell = (2\ell-1) D_{\ell-1}$, with $D_0 = 1$.
Equations \eqref{eq:struct_der_0}--\eqref{eq:struct_der_6} show that only the derivatives of order equal to or greater than $\ell$ contribute to $a_{\ell,0}$ and that the multiplicative factor in front of each derivative strongly decreases with the order of the derivative. Since the different frequency dependencies of the derivatives of different orders, this property and the typical overall scaling of $a_{\ell,0} (\nu, \beta)$ almost proportional to $\beta^{\ell \cdot p}$ mentioned in Sect. \ref{sec:theoframe} do not imply that at each multipole $\ell$ the terms from the derivatives of order greater than $\ell$ are in general not relevant.

\subsection{Relevant scalings}

From Eq.\,\eqref{eq:eta_boost}, since $\nu' = \nu (1 - \beta w)/(1 - \beta^2)^{1/2}$ and omitting for easiness the suffix `BB/dist', in the Leibniz notation we have

\begin{equation}\label{eq:linkder1}
\frac{dT_{\rm th}} {dw} = \frac{dT_{\rm th}} {d\nu'} \frac{d\nu'} {dw} = \frac{dT_{\rm th}} {d\nu'} \frac{-\beta \nu} {(1 - \beta^2)^{1/2}}\, 
\end{equation}

\noindent for the first derivative, and

\begin{align}\label{eq:linkder2}
\frac{dT_{\rm th}^2} {dw^2} & = \frac{d} {dw} \left({ \frac{dT_{\rm th}} {dw} }\right) = 
\frac{-\beta \nu} {(1 - \beta^2)^{1/2}} \left[{ \frac{d} {d\nu'} \left({ \frac{dT_{\rm th}} {d\nu'} }\right) }\right] \frac{d\nu'} {dw} 
\\& = \frac{dT_{\rm th}^2} {d\nu'^2} \left[{ \frac{-\beta \nu} {(1 - \beta^2)^{1/2}} }\right]^2\, \nonumber
\end{align}

\noindent for the second one.

In general, being the factor ${-\beta \nu} / {(1 - \beta^2)^{1/2}}$ independent of $w$, for the subsequent $n$-th derivatives we have\cite{2024AA...684A..82T}

\begin{equation}\label{eq:linkdern}
\frac{dT_{\rm th}^n} {dw^n} = \frac{dT_{\rm th}^n} {d\nu'^n} \left[{ \frac{-\beta \nu} {(1 - \beta^2)^{1/2}} }\right]^n\, .
\end{equation}

From Eqs. \eqref{eq:linkder1}--\eqref{eq:linkdern}, the derivatives $T_{\rm th}^\dn$ can be rewritten setting  

\begin{equation}\label{eq:der_at_w0}
\frac{dT_{\rm th}^n} {d\nu'^n} \rightarrow {\frac{dT_{\rm th}^n} {d\nu'^n}}\Bigg|_{w=0} = {\frac{dT_{\rm th}^n} {d\nu'^n}}\Bigg|_{\nu'_{\beta,\perp}}\, ,
\end{equation}

\noindent where $\nu'_{\beta,\perp} = \nu /(1 - \beta^2)^{1/2}$ is estimated at $w=0$ (or $\theta=\pi/2$).

Assuming a speed, $\beta_a$, different from $\beta$, and evaluating at these speeds the ratio between the derivatives $T_{\rm th}^\dn$ we obtain

\begin{equation}\label{eq:ratio_der}
\frac{{T_{\rm th}^\dn}\Bigr|_{\beta}} {{T_{\rm th}^\dn}\Bigr|_{\beta_a}}  =  f_a^{-n} \, \left( { \frac{1-\beta_a^2}{1-\beta^2} } \right)^{n/2} \, R_n \, ,
\end{equation}

\noindent where $f_a = \beta_a /\beta$ and, defining $\nu'_{\beta_a,\perp}$ for $\beta = \beta_a$,

\begin{equation}\label{eq:R}
 R_n =  \Bigg ( {\frac{dT_{\rm th}^n} {d\nu'^n}}\Bigg|_{\nu'_{\beta,\perp}} \Bigg ) \, \Bigg ( {\frac{dT_{\rm th}^n} {d\nu'^n}}\Bigg|_{\nu'_{\beta_a,\perp}} \Bigg )^{-1}.
\end{equation}

For both the speeds significantly less than unity, we can set $\nu'_{\beta_a,\perp} \simeq \nu'_{\beta,\perp}$.
All the three factors in the right hand side of Eq. \eqref{eq:ratio_der} are in principle different from unity. On the other hand, except for possible functions $T_{\rm th}$ with extreme variation in frequency, $ f_a^{-n}$ is the only term that can be remarkably different from unity.
The extremely accurate computation of $R_n$, i.e. of its very little difference from unit, would require an analogous knowledge of the change of the corresponding order derivative in an extremely narrow range between ${\nu'_{\beta,\perp}}$ and ${\nu'_{\beta_a,\perp}}$, which typically is the missing information for tabulated functions, calling for the methods discussed in what follows. Setting $R_n$ = 1 do not significantly affect the results.

\subsection{Interpolation versus derivatives}
\label{sec:InterpVSderiv}

When considering a monopole frequency spectrum characterized by a tabulated function, the estimation of the quantities $T_{\rm th}^{\rm BB/dist} (w)$ and $T_{\rm th}^\dn$ needs the interpolation of the corresponding functions in the adopted grid of points.\footnote{When adding the CMB BB monopole spectrum to the considered background, the sum should be performed in terms of additive quantities such as the antenna temperature.} Refs. \citenum{1988MC} and \citenum{1998SIAMR..40..685F} provided the weights for the (possibly centred) approximations at a particular point for the generation of finite difference formulas on arbitrarily spaced grids for any order of derivative. 
In the differentiation scheme, for a certain number of grid points, $n_{\rm ds}$, the order of accuracy decreases with the order of derivative. We explored the possibility of increasing the value of $n_{\rm ds}$ or fixing the order of accuracy while varying $n_{\rm ds}$ according to the order of derivative. Both the approaches resulted in almost the same results as long as the fourth order of accuracy at the highest derivative is achieved. For simplicity, we fixed $n_{\rm ds} = 9$ achieving the fourth order of accuracy up to the derivative of order six. For each value of $\nu'_{\beta,\perp}$, we selected 4 points on the left and 4 on the right, to keep an almost centred approximation.
Thus, the $a_{\ell,0}$ coefficients can be directly computed adopting the weights in Ref.\,\citenum{1988MC} for the zero order derivative when using Eqs.\,\ref{eq:struct_even} and \ref{eq:struct_odd} ({\it interpolation scheme}) or Eqs.\,\ref{eq:struct_der_even} and \ref{eq:struct_der_odd} ({\it derivative scheme}).

%%%%%%%%%%%%%%%%%
\section{Filtering}
\label{sec:filt}

In case of tabulated monopole frequency spectra, the presence of inaccuracies can preclude an accurate computation of the boosting effects.
One solution is to apply a pre-filtering to the monopole spectrum. It can be used independently or in combination with subsequent filtering techniques.

\subsection{Pre-filtering of monopole spectrum}
\label{sec:prefilt}

The pre-filtering of the tabulated monopole frequency spectrum smooths out inaccuracies appearing at small scales in a suitable equispaced real space variable, $u$. 
Assuming a low-pass Gaussian filtering in the Fourier space, we computed the monopole spectrum fast Fourier transform (FFT) tabulated in a grid $u_i$ ($i=1,N_p$, with $N_p$ a power of 2), then smooth it out at the modes, $f_i = 1, N_p$, corresponding to the small scales in the real space, and finally derive a filtered monopole spectrum applying the inverse FFT (FFT$^{-1}$) to the smoothed FFT. In this approach, the level of smoothing in Fourier space is given by the parameter $\sigma_f$. We set $\sigma_f  = f N_p$, 
where the value of $f$ should be properly chosen. Indeed, larger (smaller) values of $f$ translates into negligible (excessive) smoothing, possibly affecting significantly the original background shape.\cite{2024AA...684A..82T}

%Once derived the smoothed version of the original tabulation, the calculation of the differences $\Delta a_{\ell,0}$ and $\Delta R$ has been performed.

\subsection{Filtering of derivatives in real or Fourier space}
\label{sec:filt_der}

Another suitable approach, that can be also applied after the pre-filtering, is filtering the derivatives computed with the centred approximation scheme in Ref.\citenum{1988MC}. 

For each $u_i$ point in the grid, we initially compute the first derivative at all the $n_{\rm ds}$ contiguous points, $u_j$, then apply a low-pass Gaussian filter in real space to smooth out the first derivative in the $u_i$ point, evaluated as an average over the $n_{\rm ds}$ points $u_j$ with weights

\begin{equation}\label{eq:GRfilt}
G_{ij} \propto {\rm e}^{-\frac{1}{2}\left({\frac{u_j-u_i}{\sigma_G}}\right)^2} \, ,
\end{equation}

\noindent where $\sigma_G$ establishes the level of smoothing, increasing with $\sigma_G$. 

Then, the second derivative for each point is evaluated as the derivative of the filtered first derivative, and again smooth it out, iterating this scheme up to the sixth order derivative. 
In doing so, just the weights for the derivative of orders zero and one are used.\footnote{It is possible to choose different $\sigma_G$ values at each iteration in order to optimise the filter according to the amplitude of the fluctuations occurring at a given derivation step. Since this option does not provide a clear significant benefit, the $\sigma_G$ value was fixed for all the steps.} As an alternative, it is possible to filter the derivatives in the Fourier space, as discussed above. 

\subsection{Boosting amplification and deamplification}

In general, to appreciate the spectral shape details, it is not feasible to degrade too much the adopted grid resolution, particularly in the case of feature-rich spectral shapes. Instead, allowing for a hypothetical observer having a much higher speed of a significant factor $f_a$ with respect to the real speed, the Doppler shift would be larger. In this case, the boosting effects are amplified with respect to the real case, reducing the relative impact of numerical uncertainties in the spectrum characterization.

Thus, we evaluated the spherical harmonic coefficients adopting $\beta_a = f_a \beta$ while keeping $\beta_a < 1$, and then rescaled the results to the real $\beta$ value. For each observational frequency, $\nu$, we worked locally keeping $f_{a}$ as small as possible according to the tabulation grid without moving outside the range of relevant frequencies.\footnote{For an equispaced grid in log\,$\nu$ the value of $f_a$ is constant for all the grid points while for an equispaced grid in $1+z \propto 1/\nu$, $f_a$ increases with $\nu$. See Sect.\,\ref{sec:tabfunc}.} In general, the value of $f_a$ increases for decreasing resolution (or increasing step).

We used both the interpolation and the derivative scheme, and rescaled the results with Eq.\,\eqref{eq:ratio_der}. In the former case, this rescaling can be globally performed applying it as in the case of the leading derivative, of order equal to $\ell$, while, in the latter case, this rescaling can be performed for each derivative order.

\section{Results}
\label{sec:res}

In this section, I present some predictions based on the formalism described.

\subsection{Background spectra characterized by analytical functions}

Various early processes or unconventional heating sources, possibly arisen before the end of the phase of matter-radiation kinetic equilibrium, yield to Bose-Einstein (BE) like distortions, characterized by a frequency dependent chemical potential $\mu(x)$, because of the combined action of Compton scattering and photon production processes (radiative or double Compton scattering and bremsstrahlung). Accounting only for Compton scattering, the stationary solution of the standard Kompaneets equation reduces to a BE photon distribution function\citep{1970Ap&SS...7...20S} with a frequency independent, chemical potential, which also well approximates the BE-like distortion at high frequency. For mechanisms that nearly conserve the photon number density, $\mu$ turns to be proportional to the fractional energy exchanged in the plasma during the interaction, $\Delta \varepsilon/ \varepsilon_{\rm i}$, the subscript denoting the process initial time.

In addition, various sources of photon and energy injections in cosmic plasma generates Comptonization distortions,\cite{1972JETP...35..643Z} via electron heating (or cooling), and free-free (FF) distortions, because of matter ionization.
These distortions can be generated both before \cite{2012MNRAS.419.1294C} and after the cosmological recombination epoch, \cite{1986ApJ...300....1S, 1994LNP...429...28D} an unavoidable source of these signatures being the cosmological reionization associated to the various formation phases of bound structures. The Comptonization parameter, $u$, proportional to the global fractional energy exchange between matter and radiation in the cosmic plasma, and the FF distortion parameter, $y_{B} (x)$, defined by an integral over the relevant redshift interval, are the key parameters, strictly coupled, that characterize the amplitude of these imprints.

\begin{figure}
    \hskip -3mm
        \includegraphics[width=13.cm]{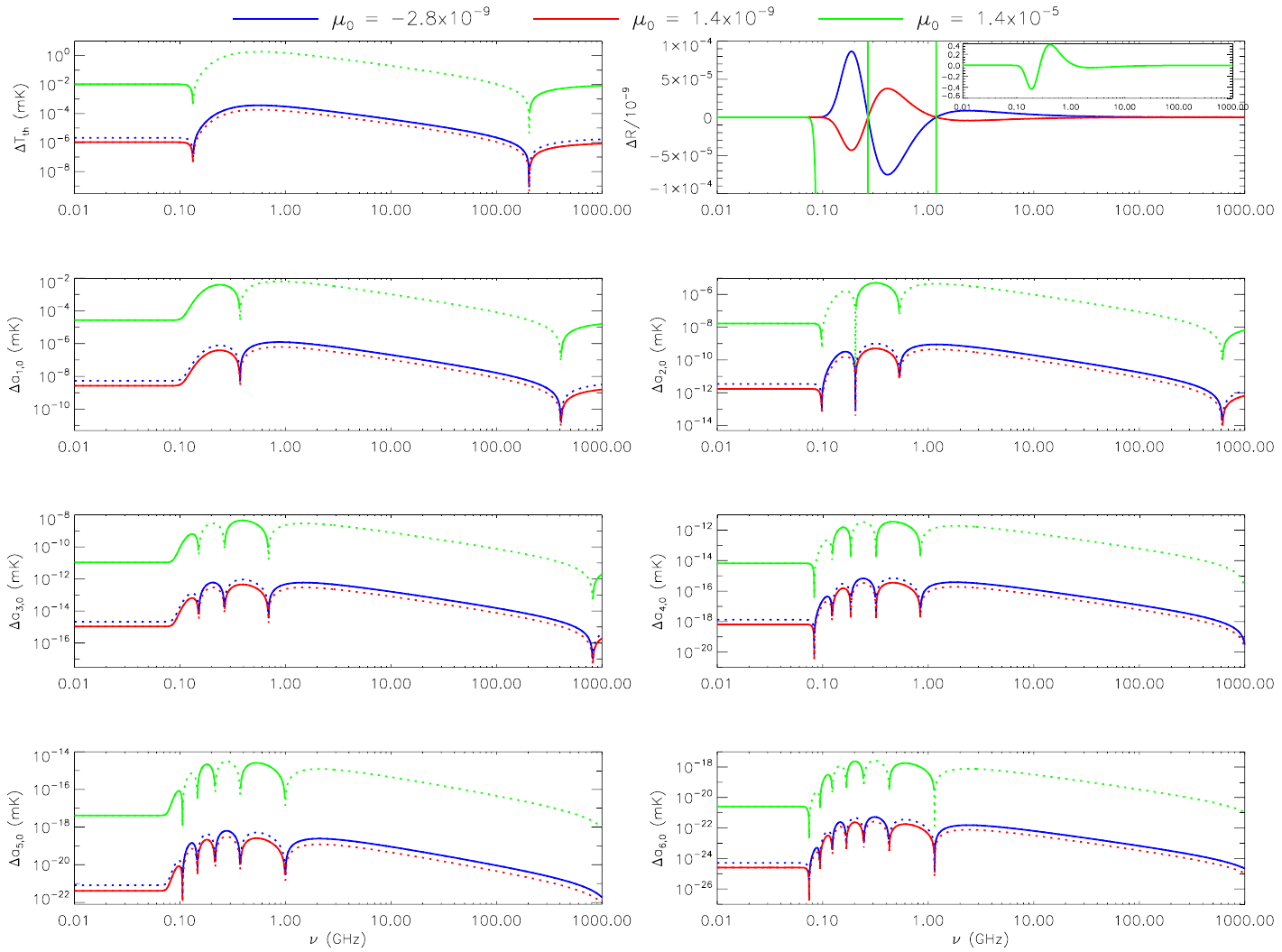}
    \caption{$\Delta T_{th}$, $\Delta R$ and $\Delta a_{\ell,0}$ for different values of the chemical potentials $\mu_{0}$ referred to the end of the kinetic equilibrium phase. Blue: from almost standard adiabatic cooling; red: as expected in almost minimal models of primordial perturbation dissipation; green: below FIRAS limit as from early decay of massive particles with suitable parameters. From Trombetti et al., A\&A {\bf 646}, p. A75, 2021 (Ref. \citenum{2021A&A...646A..75T}), reproduced with permission from Astronomy \& Astrophysics, \copyright\ ESO.}
    \label{fig:BE}
\end{figure}

\begin{figure}
    \hskip -3mm
        \includegraphics[width=13.cm]{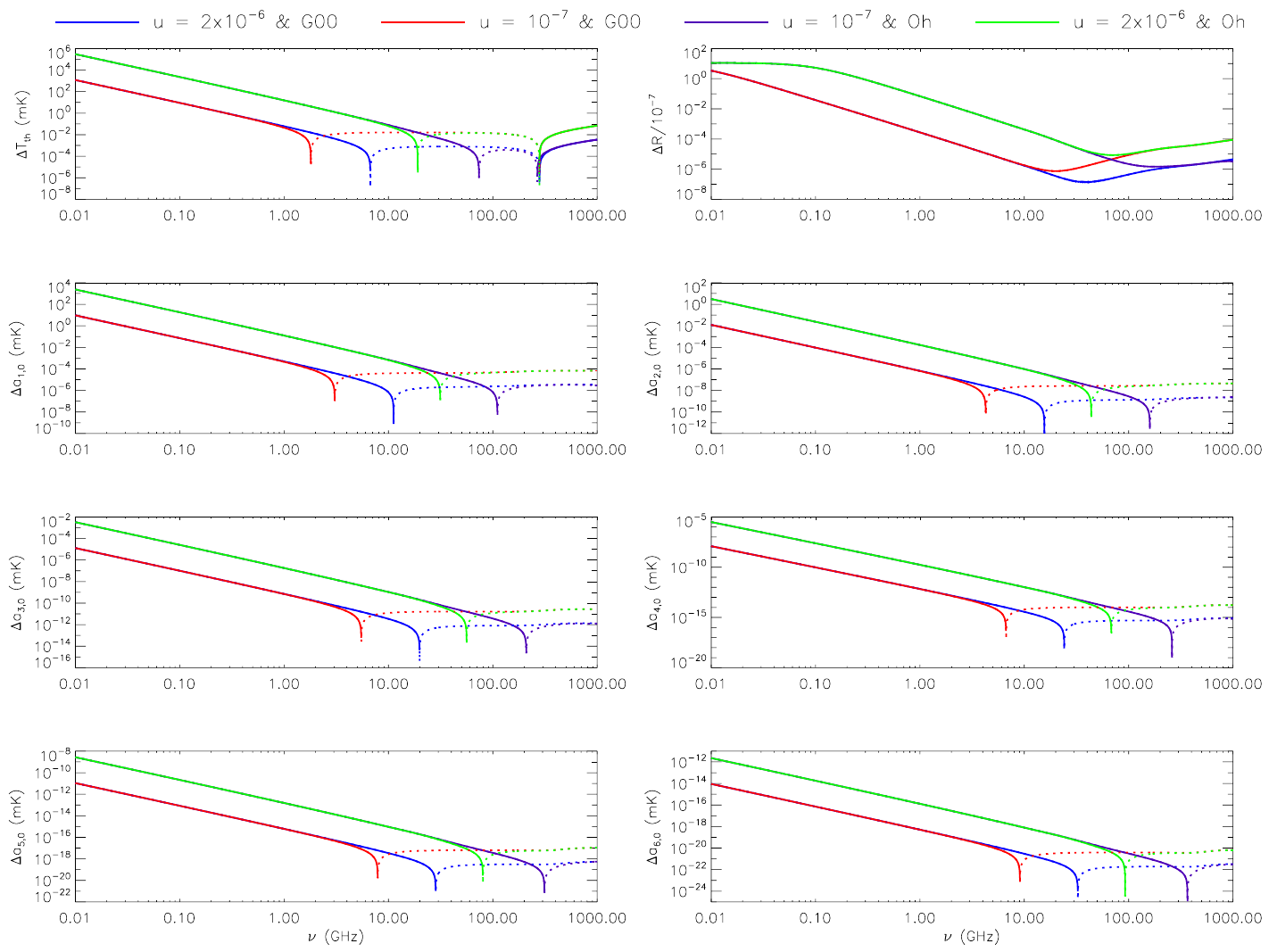}
    \caption{$\Delta T_{th}$, $\Delta R$ and $\Delta a_{\ell,0}$ for the considered combined Comptonization and diffuse FF distortion models. Solid lines (or dots) correspond to positive (or negative) values. G00 refers to the model by Ref.\citenum{2000ApJ...542..535G}, while Oh refers to the halo model in Ref.\citenum{1999ApJ...527...16O}. From Trombetti et al., A\&A {\bf 646}, p. A75, 2021 (Ref. \citenum{2021A&A...646A..75T}), reproduced with permission from Astronomy \& Astrophysics, \copyright\ ESO.}
    \label{fig:FFeC}
    \end{figure}

For analytical or semi-analytical functions, as for the BE-like or the Comptonization plus FF distortions, it is possible to explicit compute the solutions up to $\ell_{max}$. Figs.\,\ref{fig:BE} and \ref{fig:FFeC} report, in equivalent thermodynamic temperature, the difference between the intrinsic monopole spectrum and the present CMB BB spectrum (top left panel); the ratio, $R = (a_{0,0} (\nu,\beta) / \sqrt{4\pi}) / T_{th} (\nu)$, between the equivalent thermodynamic temperature of observed and intrinsic monopole, expressed as the difference $\Delta R = R - R^{\rm BB}$, where $R^{\rm BB} \simeq (1 - 2.5362 \times 10^{-7})$ is the same ratio but for the CMB BB case (top right panel); the spherical harmonic coefficients differences (other panels).\cite{2021A&A...646A..75T} 

In particular, from Fig.\,\ref{fig:BE} emerges a spectral shape opposite in sign when the parameter $\mu_{0}$ is negative, as in the case of the faster decrease of the matter temperature relative to the radiation temperature in the expanding Universe, a plateau below 0.1 GHz and an enrichment of the features at each increasing multipole. The difference $\Delta R$, displayed in top right panel, is frequency dependent and is characterized by a symmetric shape with respect to the sign of the chemical potential.

For $\ell \ge 1$, Fig.\,\ref{fig:FFeC} shows positive (negative) differences at low (high) frequencies where the FF (Comptonization) dominates. The frequencies at which the transition occurs range between 3 GHz and about 300 GHz depending on the relative amplitude of $y_B$ and $u$, and increase with the multipole. As before, top right panel reveals the frequency dependence of $\Delta R$.

\subsection{Tabulated functions}
\label{sec:tabfunc}

The 21cm line corresponds to the spin--flip transition of neutral hydrogen (HI) to the ground state. This signal is represented in terms of an offset of the 21cm brightness antenna temperature from the background temperature,\footnote{Typically, $T_{\rm back}$ is set to the CMB temperature but it could also account for signal combinations.} $T_{\rm back}$, along the observed line of sight at a frequency $\nu$ that, due to the cosmic expansion, is described by $\nu = \nu_{21 {\rm cm}} / (1+z)$ where $\nu_{21 {\rm cm}} = c / (21 {\rm cm})$ is the rest frame frequency. Thus, since at each frequency corresponds a specific redshift, the 21cm line provides a tomographic view of the cosmic evolution. 

Experimentally, observing and well characterizing the redshifted 21cm line from the diffuse HI is challenging, also because of the presence of the much stronger foreground signals.
From the data of the Experiment to Detect the Global EoR Signature (EDGES), Ref.\citenum{EDGESobs2018Nature} found a remarkable absorption profile centred at $(78 \pm 1)$ MHz, represented in terms of a flattened Gaussian identified by a set of best-fit parameters. Starting from the analytical formulation of this signal, we evaluated the monopole spectrum at each 451 frequency points in a range between $\simeq$\,28 GHz and $\simeq$\,240 GHz, and added a random Gaussian simulated noise with an amplitude relative to the signal $r_{err} = 10^{-3}$, able to represent potential inaccuracies affecting the functional representation of the background.

\begin{figure}[htb]
\centering
    \hskip -1.5mm
    \begin{minipage}[t]{0.49\linewidth}
        \includegraphics[width=6.55cm]{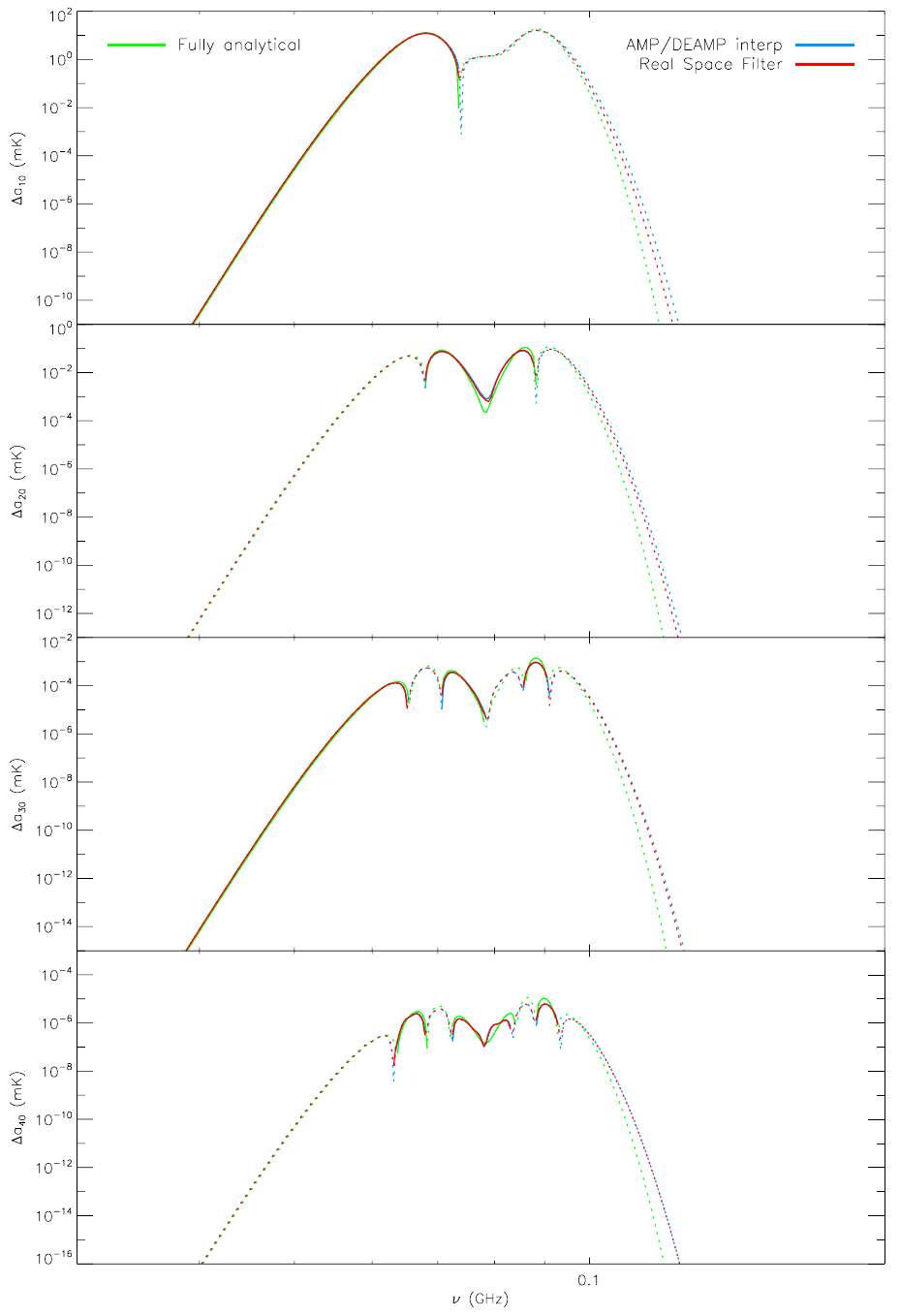}
    \end{minipage}
    \hskip 1mm
    \begin{minipage}[t]{0.49\linewidth}
        \includegraphics[width=6.55cm]{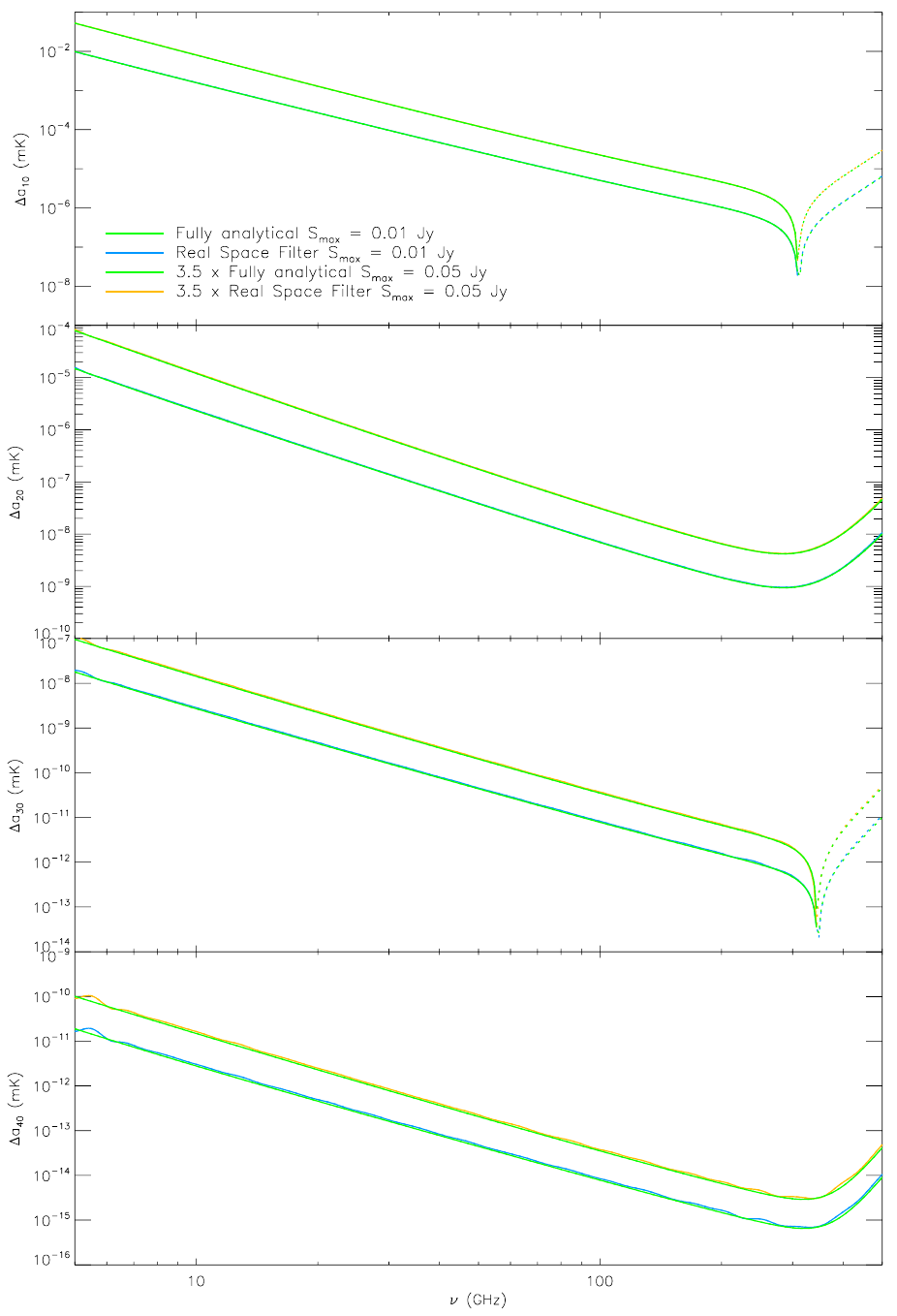}
    \end{minipage}  
    \caption{Spherical harmonic coefficients for the EDGES absorption profile (left panel), applying the pre-filtering and the two best filtering methods: the real space filter and the boosting amplification and deamplification in the interpolation scheme, and for the ESMB tabulated intensity (right panel) derived adopting two different $S_{max}$ thresholds and the corresponding ideal fully analytical case. See legend and text. Adapted from Trombetti et al., A\&A {\bf 684}, p. A82, 2024 (Ref. \citenum{2024AA...684A..82T}), reproduced with permission from Astronomy \& Astrophysics, \copyright\ ESO.}
    \label{fig:21cmESMB}
\end{figure}

Left panel of Fig.\,\ref{fig:21cmESMB} shows the spectra up to $\ell = 4$ once the pre-filtering and a Gaussian real space filter or the amplification and deamplification technique have been applied in the interpolation scheme. At very low $\ell$'s, it emerges the (already discussed in Sect.\,\ref{sec:prefilt}) smoothing excess around the relative minima and maxima due to the pre-filtering. For smaller $r_{err}$ values, this effect is strongly alleviated by avoiding the pre-filtering. 

At radio frequencies, AGN and star-forming galaxies provide similar contribution to the background,\cite{2023MNRAS.521..332T} while at higher frequencies, sources that prevail in the extragalactic and microwave sky are fuelled by AGN. Their observed flux density, indeed, comes from synchrotron radiation due to accelerated relativistic charged particles. Radio AGN frequency spectra are typically characterized by a power law with $S\propto\nu^{\alpha}$, with $\alpha<-0.5$ for steep spectrum sources or $-0.5<\alpha<0.5$ for flat ones. These spectra can emerge in extended radio lobes (the former) or compact regions of radio jets (the latter).

The \lq AGN unified model\rq\,\citep{urr95,net15} assumes that the main difference between the two spectra is given by a different orientation of the observer line of sight and the axis of the specific jet emerging from the central black hole. A side-on view of the jet-axis translates into a steep spectrum, while for a line of sight close to the axis a flat spectrum compact source is observed, the blazar.\citep{dez10} Thus, the most abundant population at radio frequencies ($\nu < 10-20$ GHz) is the steep spectrum one. On the other hand, it becomes sub-dominant at higher frequencies owing to its spectrum steepness, and flat spectrum sources dominate from few tens of GHz.

I considered the extragalactic radio sources model of Ref.\,\citenum{lag20}, an updated version of Ref.\,\citenum{tuc11}, where the spectrum of flat sources is expected to break at some frequency between $10$ and $1000$ GHz and to steepen at higher frequencies, due to the transition of the observed synchrotron emission from the optically thick to the optically thin regime and to electrons cooling effects. In particular, flat-spectrum radio quasars (FSRQs) have a break frequency $\nu < 100$ GHz, while in BL-Lacs it should appear at $\nu \gsim100$ GHz.

From the differential number counts of the model, we derived the extragalactic background intensity between 1\,GHz and 1\,THz in a grid of 512 frequencies equispaced in logarithmic units down to $S_{min}\sim10\,\mu$Jy and five threshold values of $S_{max}$, from 0.01 to 0.1\,Jy. Then, we computed the spherical harmonic coefficients for the Extragalactic Source Microwave Background (ESMB) tabular representations under study. As expected, the quality of the predictions does not vary much with the assumed threshold, thus, for simplicity, we report the results only for two different values of $S_{max}$. Right panel of Fig. \ref{fig:21cmESMB} shows the spectra up to $\ell = 4$ for $S_{max} = 0.01$ Jy and 0.05 Jy (multiplied by a 3.5 factor to better distinguish the two curves). Here, a pre-filtering is applied together with the Gaussian filter in real space. Except for some little oscillations at the lowest frequencies for $\ell = 4$, this filtering approach has been found to well reproduce the expected features.

\section{Observational perspectives}
\label{sec:obsperspe}
In the previous sections, several methods have been presented to compute the modification of the monopole spectrum and its transfer to higher multipoles when dealing with the observer peculiar motion. Furthermore, the spherical harmonic coefficients have been derived for some unavoidable spectral distortions among the many processes, or combination of them, occurred in the cosmic plasma at early or late epochs.

Assuming different foreground and calibration residuals, and a sensitivity of a Cosmic Origins Explorer (CORE) like experiment, the potential improvement with respect to FIRAS in the recovery of the distortion parameters, as the Comptonization parameter $u$ and the BE chemical potential $\mu_{0}$ has already been evaluated in Ref.\citenum{2018JCAP...04..021B}. CORE has been designed to have 2100 detectors, about $45\%$ of which are located in CMB frequency channels between 130 and 220 GHz.\cite{2018JCAP...04..014D} Those (six) CMB channels yield an aggregated CMB sensitivity of 2 $\mu$K$\cdot$arcmin. Hence, for an ideal case of perfect calibration and foreground subtraction, the improvement can be of 500--600 and 600--1000 for $u$ and $\mu_{0}$, respectively.

In general, CMB anisotropy experiments, and mainly dipole based studies, are affected by relative calibration accuracy. The presence of residual foregrounds is critical in absolute as well as differential approaches.
Even considering potential foreground relative residuals of the order of $10^{-2}$ and relative calibration uncertainties of the order of $10^{-4}$ at $\simeq$ degree scale for a full sky approach, the improvement factor with respect to FIRAS is still very promising, of $\simeq 20$ and $\simeq 40$ for $u$ and $\mu_{0}$, respectively. Better results ($\sim 20\%$) can be derived by applying sky masks in order to reduce the impact of potential residuals and, obviously, by improving foreground treatment and calibration.

In the case of almost independently observed sky patches, as in interferometric techniques, and for extreme sensitivity measurements, the dipole pattern can be reconstructed also in limited sky areas.
Indeed, even though the full amplitude of the dipole pattern clearly emerges at large angular scales, it is also possible to detect it in a small sky patch along a meridian, as evident from the zoom-in part of Fig.\,\ref{fig:mapFFeC} and described in the next section.

\subsection{Sky patches}
\label{sec:skypatch}

As anticipated, the dipole pattern amplitude reconstruction could be carried out not necessarily requiring a coherent mapping up to the largest angular scales. Indeed, along a meridian in a reference frame with $z$-axis parallel to the observer motion, the signal variation, from a dipole pattern $\Delta T$, at colatitude $\theta$ within a limited sky area of linear size $\Delta \theta$ has an amplitude $|\Delta T_{\Delta \theta}| \simeq \Delta T \cdot (\Delta \theta/90^{\circ})$ sin\,$\theta$, with sin\,$\theta$ not far from 1 at angles large enough from the poles.\cite{2024AA...684A..82T}

\begin{figure}
    \hskip -3mm
        \includegraphics[width=13.cm]{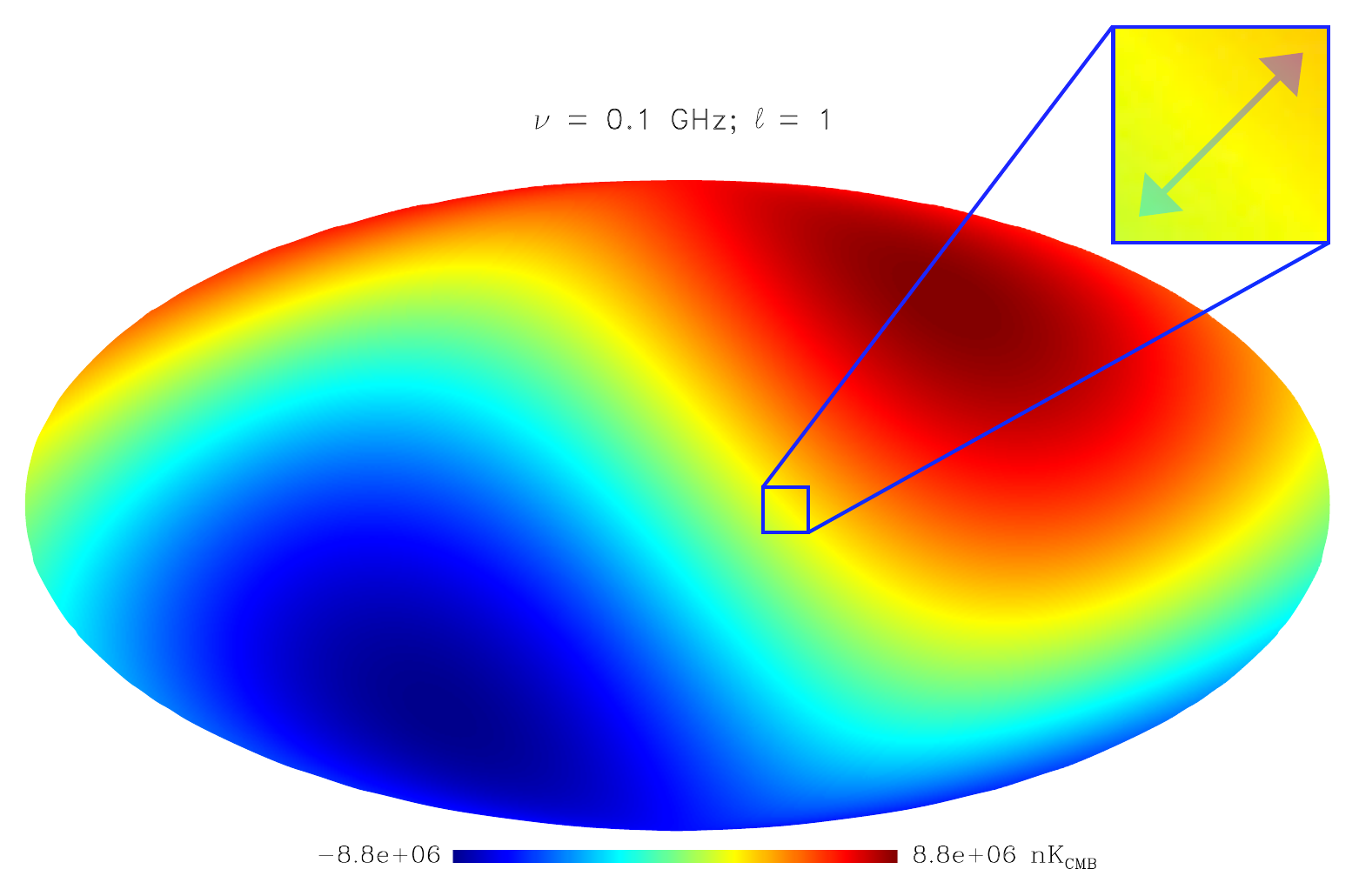}
    \caption{Dipole sky map in Galactic coordinates for a combined Comptonization and diffuse FF distortion model at 0.1\,GHz (where FF is the dominant process). According to Ref.\citenum{2020A&A...641A...1P}, we adopt $l = 264.021^\circ$ and $b = 48.253^\circ$ for the dipole direction. The zoomed inset shows that, also in a restricted sky area, it is possible to detect a dipolar modulation.}
    \label{fig:mapFFeC}
    \end{figure}

As an example, Fig.\,\ref{fig:mapFFeC} shows the full sky temperature pattern for a Comptonization distortion characterized by a parameter $u = 10^{-7}$ combined to FF distortion expected from the integrated contribution of a set of ionized halos at substantial redshift, described by a parameter $y_{B} \sim 1.5 \times 10^{-6}$ at $\nu \sim 2$ GHz, as predicted in the model by Ref. \citenum{1999ApJ...527...16O}. %Indeed, even though a dipole pattern in the sky clearly emerges, it is also possible to detect the signal in a small sky patch along a meridian, as described in Sect.\,\ref{sec:skypatch} and evident from the zoom-in part of Fig.\,\ref{fig:mapFFeC}.

\begin{figure}[htb]
\centering
    \begin{minipage}[t]{0.49\linewidth}
        \includegraphics[width=6.5cm]{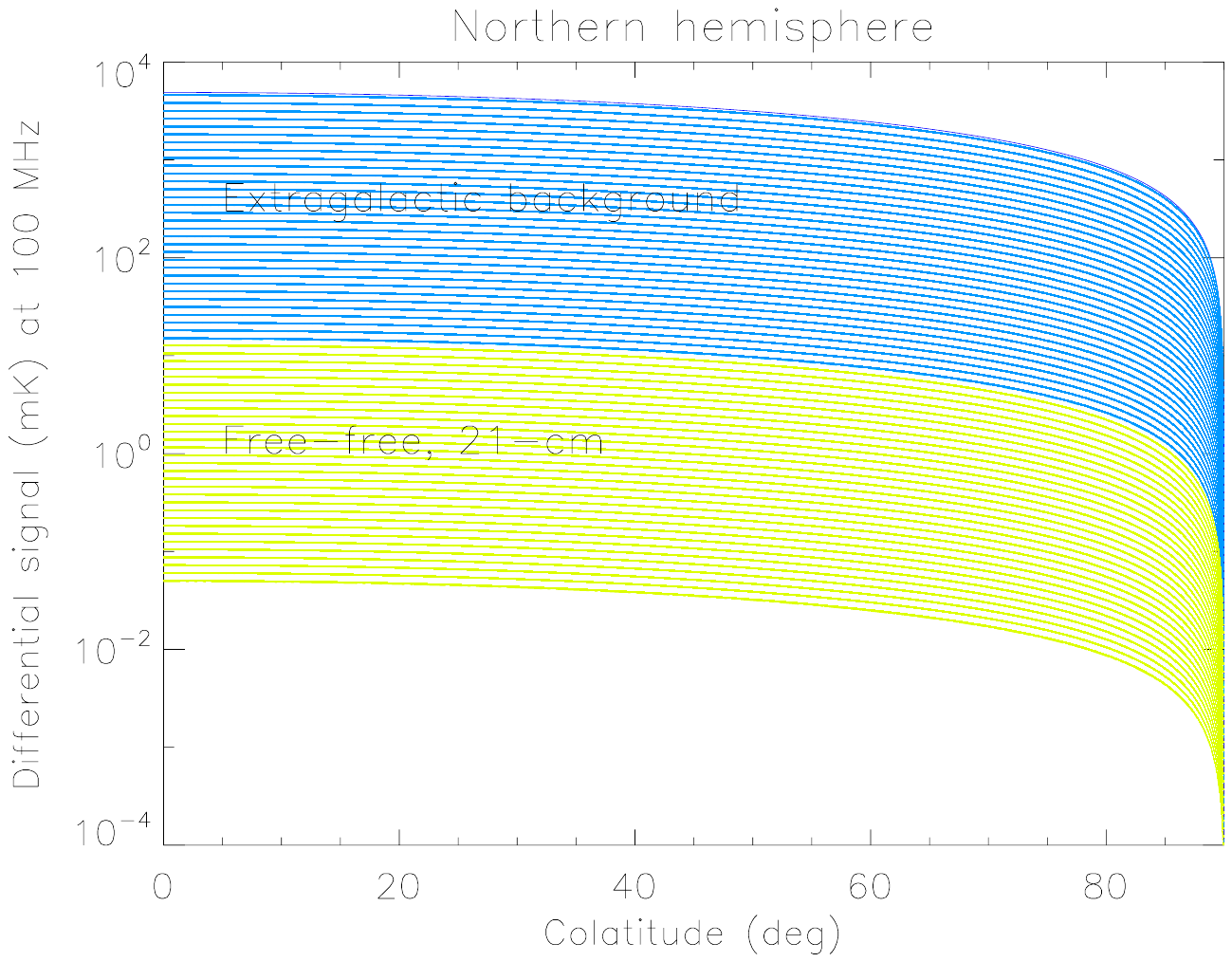}
    \end{minipage}
    \hskip 1.42mm
    \begin{minipage}[t]{0.49\linewidth}
        \includegraphics[width=6.5cm]{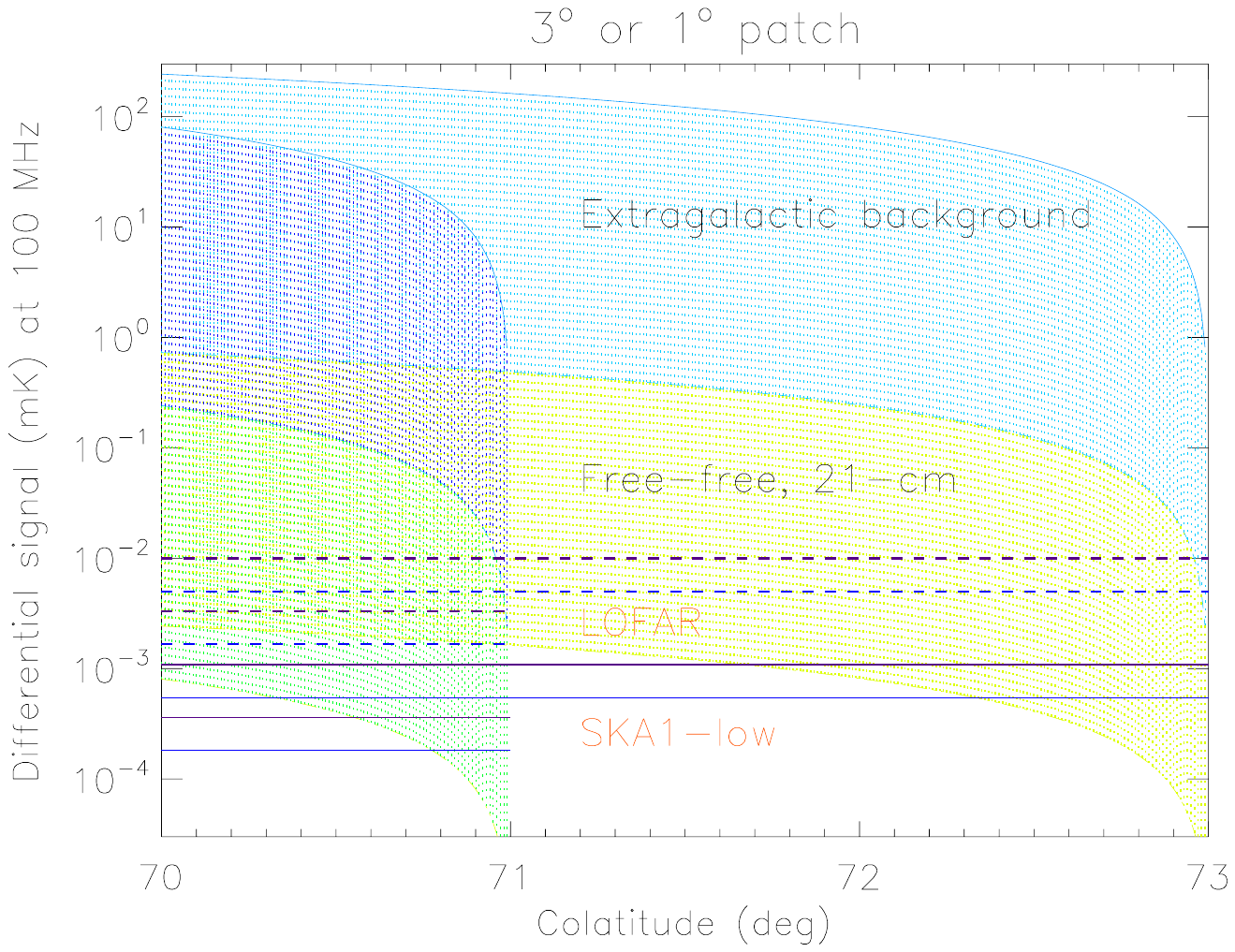}
    \end{minipage}  
    \caption{Range of predicted differential signal for the dipole, $\Delta T(\theta) = \Delta a_{\ell,0} $ $[3/(4\pi)]^{1/2} $ $ {\rm cos} \theta$, where $\theta$ is the colatitude and $\Delta a_{\ell,0}$ refers to the dipole harmonic component $\ell$\,=\,1, $m$\,=\,0 in a frame with the $z$-axis parallel to the observer motion, after the subtraction of the standard CMB BB, in order to emphasize the interesting signal (here at 100 MHz and in equivalent thermodynamic temperature). Left panel: the case of an all-sky survey, displayed for simplicity only for a hemisphere. Right panel: a zoom of left panel for a patch of 3$^{\circ}$ (1$^{\circ}$); here we display $\Delta T(\theta) - \Delta T(\theta_{*})$, with $\theta_{*}$ = 73$^{\circ}$ (71$^{\circ}$), i.e. the differential signal inside the patch, to be compared with typical sensitivity levels for LOFAR (dashed lines) and SKA1-low (solid lines) in a nominal pixel of 2 arcmin for one day of integration in the patch. Violet (blue) lines assumes a bandwidth of 10 (40) MHz to appreciate spectral shapes of the 21-cm redshifted HI line (the other types of signal).
    }
    \label{fig:dippatch}
\end{figure}

For instance, see left panel of Fig.\,\ref{fig:dippatch}, at $\nu \sim (50 - 100)$ MHz, we find that the relevant signals (after the subtraction of the standard CMB BB spectrum dipole) have amplitudes $ \Delta T_{\rm th} - \Delta T_{\rm th}^{\rm BB} \sim 0.1 - 50$ mK for the FF and 21cm signatures depending on the model. Similar values also hold for the radio extragalactic background when the deepest detection thresholds are considered.
Values of about 2 (or 4) orders of magnitude higher are achieved for shallower detection limits (or for the best-fit radio signal of the extragalactic background). Thus, as emerges from right panel of Fig.\,\ref{fig:dippatch}, for a patch of size $\Delta \theta \sim 3^\circ$, sensitivities to the diffuse signal of $\text{about}$ several tens of mK (or in a range from $\text{about}$ a few $\mu$K to $\text{about}$ mK) could allow identifying the extragalactic background (or the reionization imprints on the diffuse radio dipole). All these sensitivity levels can certainly be achieved based on the Square Kilometre Array (SKA) specifications.\cite{Dewdney2015} In the figure, the higher part of the blue areas refers to estimates of the signals for the diffuse background from extragalactic radiosources (see Ref.\citenum{2021A&A...646A..75T}); in the lower part it is assumed that sources above certain detection thresholds are subtracted. Yellow areas refer to estimates of the signals for the diffuse FF distortion, from the minimal prediction accounting for the diffuse IGM contribution to the maximum level corresponding to the integrated contribution from ionized halos, and for various models of the IGM 21cm redshifted HI line.\cite{2019A&A...631A..61T} 

Subtracting sources by applying a broad set of detection thresholds would also help to clarify to what extent the radio extragalactic background could be ascribed to extragalactic sources or if it is a significant part of intrinsic cosmological or diffuse origin. This would contribute to answering the question about the level and origin of the radio extragalactic background, which is still controversially discussed (see Refs.\,\citenum{Subrahmanyan:2013,2018Natur.564E..32H,2018MNRAS.481L...6S}). Clearly, collecting many patches would improve the statistical information. 

\section{Conclusions}

Boosting effects can arise in the monopole frequency spectrum and transferred to higher multipoles when accounting for an observer peculiar motion. A study to evaluate the impact of this effect has been presented for some peculiar CMB spectral distortions and background spectra described by analytical or semi-analytical functions, or by tabular representations. Because of the presence of numerical uncertainties, whose effect typically increases with the derivative order, the limited accuracy in the intrinsic monopole spectrum can prevent a precise computation of the multipole pattern spectra, even for the observed monopole. To by-pass this problem, a pre-filtering method and various filtering techniques have been discussed, aimed at deriving reliable predictions for a wide range of background models up to a multipole $\ell_{max} = 6$. These approaches have been tested on analytical approximations derived from the adopted tabulation grids and contaminated by simulated noise. These methods have been applied to different types of signals of both cosmological and astrophysical interest.

Finally, observational perspectives regarding the improvement in the recovery of characteristic distortions parameters with respect to FIRAS and the reconstruction of several classes of background spectra have been supplied, considering a (typical) full sky approach as well as sky patches of a few degrees, where the dipole pattern can still be reconstructed in the case of experiments with extremely high sensitivity and resolution.

%\section*{Acknowledgments}

\bibliographystyle{ws-procs961x669}
\bibliography{biblio}

\end{document}